\documentclass[aps,prl,reprint]{revtex4-1}

\usepackage{graphicx} 
\usepackage{ae,aecompl} 
\usepackage{bm} 
\usepackage{amsmath} 
\usepackage{amssymb} 
\usepackage{textcomp} 
\usepackage{xcolor} 
\usepackage{soul}

\newcommand{\Co}{\ensuremath{\text{\,c}_{\gamma}}}
\newcommand{\Si}{\ensuremath{\text{\,s}_{\gamma}}}
\newcommand{\Ta}{\ensuremath{\text{\,t}_{\gamma}}}

\begin{document}

\title{Atomic homodyne detection of continuous variable entangled twin-atom states.}

\author{C. Gross}
\author{H. Strobel}
\author{E. Nicklas}
\author{T. Zibold}
\affiliation{Kirchhoff-Institut f\"ur Physik, Universit\"at Heidelberg,
Im Neuenheimer Feld 227, 69120 Heidelberg, Germany}
\author{N. Bar-Gill}
\altaffiliation{present address: Harvard-Smithsonian CfA, Harvard University Dept. of Physics, Cambridge, MA 02138, USA}
\author{G. Kurizki}
\affiliation{Weizmann Institute of Science, Rehovot 76100, Israel}
\author{M.K. Oberthaler}
\affiliation{Kirchhoff-Institut f\"ur Physik, Universit\"at Heidelberg,
Im Neuenheimer Feld 227, 69120 Heidelberg, Germany}

\begin{abstract}
Historically, the completeness of  quantum theory has been questioned using the concept of bipartite continuous variable entanglement~\cite{Einstein:1935}. 
The non-classical correlations (entanglement) between the two subsystems imply that the observables of one subsystem  are determined by the measurement choice on the other, regardless of their distance.
Nowadays, continuous variable entanglement is regarded as an essential resource allowing for quantum enhanced measurement resolution~\cite{Giovannetti:2004}, the realization of quantum teleportation~\cite{Braunstein:2005, Braunstein:1998, Opatrny:2001} and quantum memories~\cite{Braunstein:2005, Hammerer:2010}, or the demonstration of the Einstein-Podolsky-Rosen paradox~\cite{Einstein:1935, Reid:2009, Ou:1992, Reid:1989fk}. 
These applications rely on techniques to manipulate and detect  coherences of quantum fields, the quadratures.
While in optics coherent homodyne detection~\cite{Walls:2008} of quadratures is a standard technique,  for massive particles a corresponding method was missing. 
Here we report on the realization of an atomic analog to homodyne detection for the measurement of matter-wave quadratures. 
The application of this technique to a quantum state produced by spin-changing collisions in a Bose-Einstein condensate~\cite{Duan:2000, Pu:2000} reveals continuous variable entanglement, as well as the twin-atom character of the state~\cite{ Raymer:2003}. 
With that we present a new system in which continuous variable entanglement of massive particles is demonstrated~\cite{Julsgaard:2001, Hammerer:2010}.
The direct detection of atomic  quadratures has applications not only in experimental quantum atom optics but also for the measurement of fields in many-body systems of massive particles~\cite{Bloch:2008}. 
\end{abstract}

\maketitle

\begin{figure*}[t]
	\begin{center}
		\includegraphics{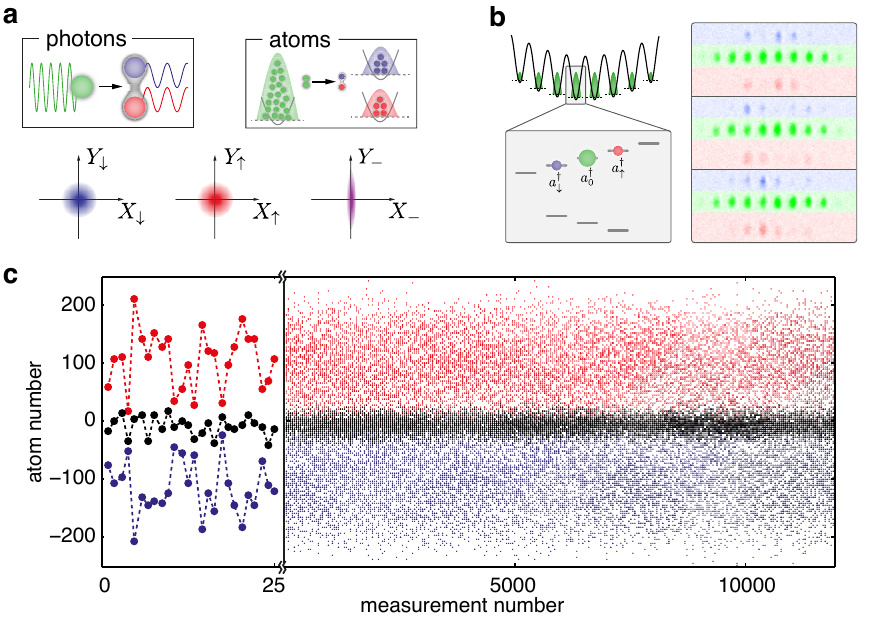} \caption{ {\bf Analogy to optics and measured population correlations of twin-atom states. a, } Parametric downconversion with light and the analogy to atomic spin changing collisions. In a nonlinear medium effective interactions result in the creation of photon pairs in the signal (red) and idler (blue) modes from pump mode photons (green). The quadratures of the output modes are centered around the origin -- the individual ones are isotropic, while the two-mode quadratures are squeezed reflecting their correlations (purple). Pair creation due to spin changing collisions in tight traps is an analogous process in quantum atom optics. 
		{\bf b, } Experimental system and exemplary raw absorption pictures. The illustration on the left shows schematically the 1D optical lattice which is superimposed by an optical dipole trap to tightly confine the atoms in all directions. In each site selected Zeeman levels of the hyperfine spin are the only accessible degrees of freedom. The states involved in the spin changing collision process are highlighted. Three typical experimental pictures after spin changing collisions are presented on the right where the clear spatial separation of sites and states can be seen.
		{\bf c,} Population correlations.  Population correlations are visible directly on the raw data where the atom numbers in the signal and idler modes fluctuate strongly (blue, red) while their difference (black) shows small noise. The left part of the graph is a zoom into $25$ experimental realizations clearly showing that the fluctuations are common mode.} 
\label{fig1} 
	\end{center}
\end{figure*}

Continuous variable entangled states, whose inter-mode entanglement is reflected in quadrature as well as population correlations, are routinely generated by parametric downconversion in optical parametric amplifiers (OPA)~\cite{Walls:2008}. Spin-changing collisions in ultracold bosonic quantum gases provide a similar nonlinear process for matter-waves generating atom pairs with spin up and down -- twin-atoms~\cite{Leslie:2009, Klempt:2010, Chang:2004a, Schmaljohann:2004}.
Correlated photon or atom pairs in the signal $|\uparrow\rangle$ and idler $|\downarrow\rangle$ modes are created from a pump field  via nonlinear interactions described by the Hamiltonian $\hbar \alpha_0(a_{\downarrow}^\dagger a_{\uparrow}^\dagger + a_{\downarrow}a_{\uparrow})$,   where $a^{\dagger}_{k}$ is the creation operator of the respective mode $k$ and $2\pi \hbar$ is Planck's constant.
This Hamiltonian presumes a large amplitude coherent pump field, whose depletion or distortion by the signal and idler modes is negligible such that it can be treated classically. 
In optics the effective nonlinearity $\alpha_0$ is proportional to the amplitude of this pump field and the susceptibility of the medium. In the analogous regime for atoms the effective nonlinearity originates from interatomic interactions and is proportional to the population in the pump mode, that is, the Bose-Einstein condensate (BEC).
For initially empty signal and idler modes quantum fluctuations are amplified and the output is the so called two-mode squeezed vacuum~\cite{Walls:2008}. 
This state is characterized by the vanishing of the mean field amplitude $\langle E_{k} \rangle = 0$ in each of the modes, while the mean intensity is nonzero $\langle I_{k} \rangle \propto \langle E_{k}^\dagger E_{k} \rangle>0$. 
The associated two-mode entanglement is revealed by the quadratures $X_{\pm}(\varphi)= a_{\downarrow} {\rm e}^{-i \varphi} \pm a_{\uparrow}^{\dagger} {\rm e}^{i \varphi} + {\rm h.c.}$, the signature being the squeezing of their variance for appropriately chosen $\varphi$.
To access the two-mode quadratures the coherent homodyne measurement technique is employed, whereby the conjugate canonical quadratures $X_{k}=a_{k}^\dagger + a_{k}$ and $Y_{k}=i(a_{k}^\dagger - a_{k})$ are measured by mixing the quantum mode with a strong classical reference field, the local oscillator $a_{\rm LO}^\dagger \approx \sqrt{N_{\rm LO}} {\rm e}^{-i \varphi}$ with large amplitude $\sqrt{N_{\rm LO}}$. 
Experimental control of the local oscillator phase $\varphi$ provides access to the continuous distribution of single mode quadratures $X_{k}(\varphi)=\mathrm{e}^{i \varphi}a_{k}^\dagger + \mathrm{e}^{-i \varphi}a_{k}$ and their two-mode counterparts $X_{\pm}(\varphi)$ ($X_{k}$ and $Y_{k}$ are used as abbreviations for $X_{k}(0)$ and $X_{k}(\tfrac{\pi}{2})$). 
For the squeezed vacuum state the quadrature distribution of individual modes is isotropic and centered around the origin reflecting the undefined phase. The correlation of the modes leads to anisotropic and squeezed distributions of the two-mode quadratures $X_{\pm}(\varphi)$ as seen in figure~\ref{fig1}a~\cite{Caves:1985, Walls:2008}.

Here we report on the development of atomic homodyne detection and apply this technique to measure quadratures of twin-atom matter-wave fields generated by controlled spin-changing collisions in a BEC.
The experiments involve a spinor BEC of Rubidium~87 trapped in a 1D optical lattice with few hundred atoms per site and a lattice spacing of $5.5\,\mu$m (Fig.~\ref{fig1}b). A high lattice potential assures that tunneling between sites is negligible on the experimental timescale such that the condensates in the different sites are independent~\cite{Gross:2010}. 
The density in the center of the lattice sites is in the order of $2 \times 10^{14}\,\mathrm{cm}^{-3}$ resulting in a minimal spin healing length of $\xi \approx 1\,\mu$m, which is comparable to the extension of the on-site wave function (approximately $1.1\,\mu$m FWHM). The spatial degrees of freedom are therefore frozen and the dynamics happens exclusively in the hyperfine spin~\cite{Law:1998}. 
The condensate is initially prepared in the $(F,\,m_F)=(2,0)$ hyperfine state serving as the analog to the pump mode in an OPA. Making use of a combination of quadratic Zeeman shift and state dependent microwave dressing~\cite{Kronjaeger:2006,Gerbier:2006} the spin dynamics is energetically restricted to a spin $1$ subspace defined by the $0,\pm1$ Zeeman states (see supplementary information). We choose a constant evolution time under spin changing collisions of $22\,$ms in all our experiments. In the regime of small pump mode depletion spin-changing collisions can be described by an OPA type pair coupling between the Zeeman states~\cite{Duan:2000, Pu:2000, Leslie:2009, Klempt:2010} with the $+1$ and $-1$ states as signal ($\uparrow$) and idler ($\downarrow$) modes.
In optics the integrated nonlinearity is restricted by short interaction times, a limit which is surpassed in the atomic system, such that continuous variable entangled states with large population can be realized.
Since the underlying process generates pairs of atoms in the signal and idler modes -- twin-atoms -- the mean and variance of the population difference $N_- = N_\downarrow-N_\uparrow$ should ideally vanish although the total population $N_+ = N_{\downarrow} + N_{\uparrow}$ fluctuates strongly. This can be experimentally confirmed by absorption imaging of the atomic cloud~\cite{Gross:2010} where state selectivity is achieved by Stern-Gerlach separation (Fig.~\ref{fig1}b,~c). 

\begin{figure*}[t]
	\begin{center}
		\includegraphics{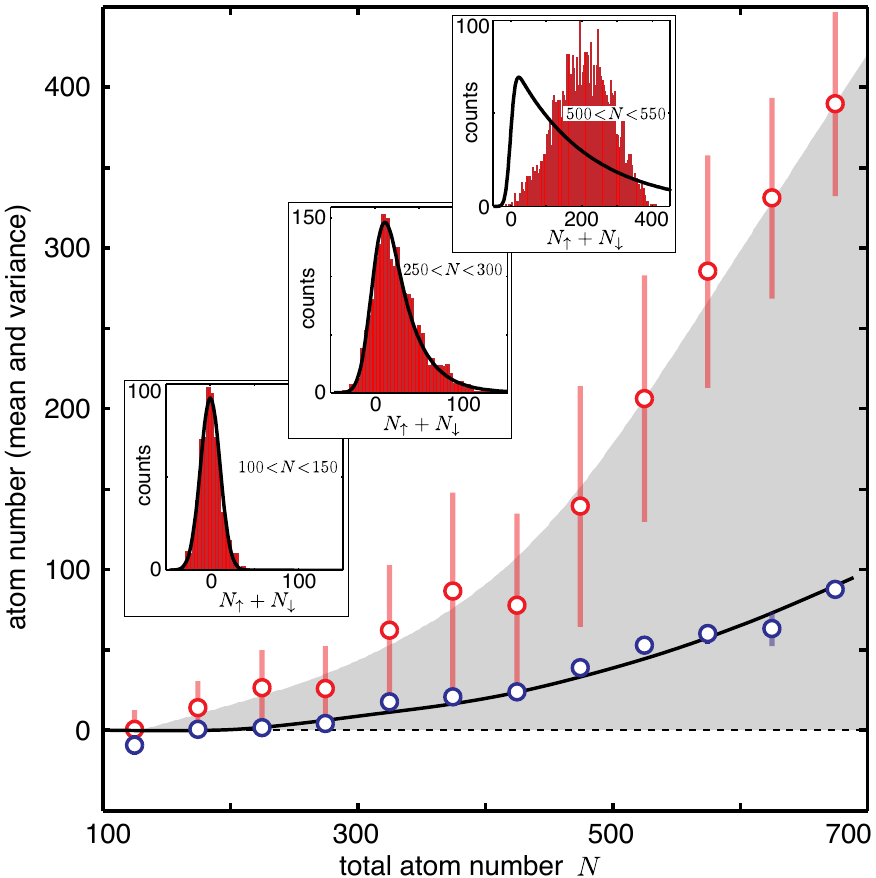} \caption{ {\bf Population of the signal and idler modes.} The red data shows the mean population $\langle N_+ \rangle= \langle N_\uparrow + N_\downarrow \rangle$ versus total atom number $N$. The bars indicate the fluctuations ($1$-s.d.) of the data. The large uncertainty is a feature of the nonlinear process itself which leads to a non-gaussian distribution of the mode population. The gray area is the sub-poisson regime and its upper bound is obtained by a spline fit to $\langle N_+ \rangle$. Blue data points show the variance of the population difference $\Delta^2 N_-$, which is compatible with zero for small $N$ (bars indicate the $1$-s.d. statistical uncertainty). Its increase with $N$ is reproduced when taking particle loss due to spin relaxation into account (black line). The insets show the distribution of the mode population for indicated total atom numbers. For small $N$ the distributions match the prediction for a non-depleted pump with the measured mean population $\langle N_+ \rangle$  (black lines) while for larger $N$ they clearly differ. The fitted squeezed vacuum distribution for $250 < N < 300$ corresponds to a squeezing parameter of $r\approx2$~\cite{Walls:2008}.} 
\label{fig2} 
	\end{center}
\end{figure*}

A quantitative analysis of the variance of the population difference $\Delta^2 N_-$ and the distribution of the population sum $N_+$ gives a first indication of the quantum state of the system (Fig.~\ref{fig2}). 
Due to the pump mode population dependent nonlinearity the mean atom number in signal and idler mode $\langle N_+\rangle$ increases nonlinearly with the total atom number $N$, resulting in a growing fraction $\langle N_+\rangle / N$. In the small $N \lesssim 300$ limit the observed distribution of the total population $N_+$ matches the prediction for the squeezed vacuum (taking detection noise into account by convolution with a gaussian) with maximal squeezing parameter $r\approx2$~\cite{Walls:2008}. The corresponding mean population is $\langle N_+ \rangle \approx 25 $ and the data shows a comparably large standard deviation -- in contrast,  the noise in the population difference $N_{-}$  almost vanishes  ($\Delta N_- < 1.9$ atoms).  It is important to note that the atomic variances reported throughout this manuscript are corrected for detection noise as detailed in the supplementary information. For larger total atom numbers $N$ the pump depletion cannot be neglected any more resulting in a breakdown of the analogy to the optical parametric amplification picture and the observed distributions of $N_+$ differ from the squeezed vacuum (insets in Fig.~\ref{fig2})~\cite{Law:1998}. In this regime the well controllable spin dynamics offer prospects for the deterministic generation of non-gaussian entangled twin-atom states~\cite{Diener:2006}. 
Density dependent two-body spin relaxation loss deteriorates the perfect pair correlations such that the relative population noise grows with $N$. Averaging the results for all total atom numbers we find the noise $\Delta^2 N_-$  suppressed by  $6.9\,$dB below the poisson level $\langle N_+\rangle$.

\begin{figure*}[t]
	\begin{center}
		\includegraphics{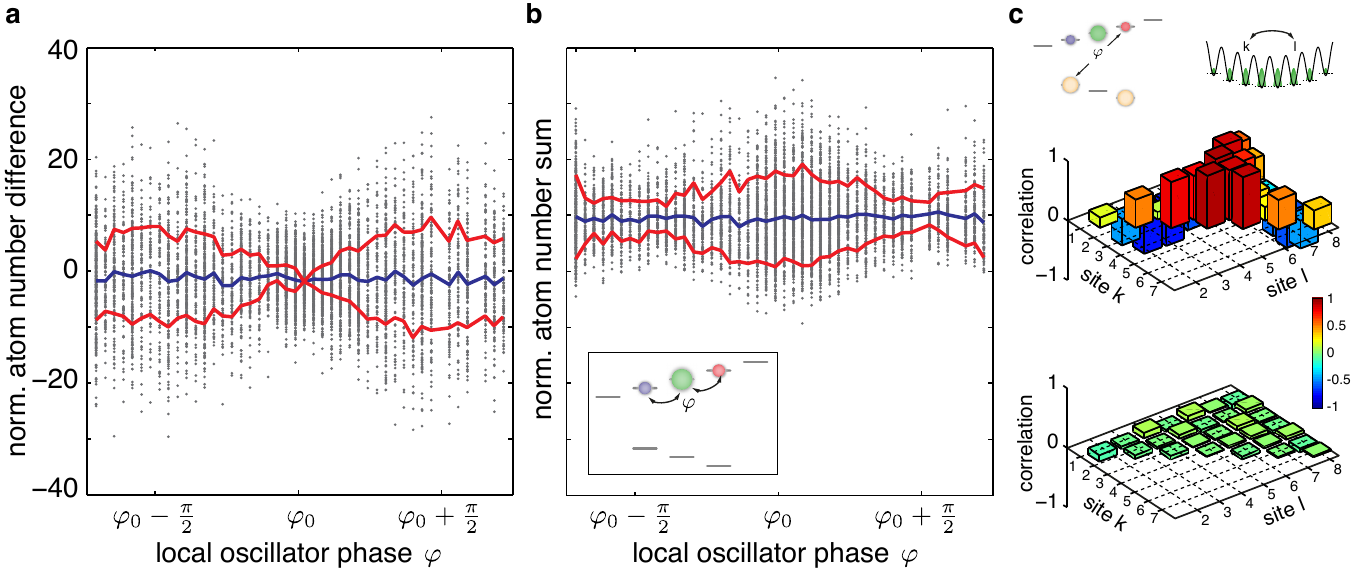} \caption{ {\bf Atomic homodyning. a, } Homodyne measurement of the quadrature difference. After symmetric coupling to the $m_F=0$ state (inset in b) the normalized population difference in the $m_F=\pm1$ states provides a measure for the two-mode quadrature difference. Phase dependent noise is visible directly on the raw data while no dependence is observed in the mean population difference (solid blue line).   
		 {\bf b, }Homodyne measurement of the quadrature sum. An upper bound for the variance of the two-mode quadrature sum is obtained from the fluctuations of the normalized population sum after the coupling. The fluctuations of the raw data are strongly phase dependent while the mean (blue solid line) stays constant. The variance minimum is shifted by $\tfrac{\pi}{2}$ as compared to a indicating noise suppression in the  orthogonal two-mode quadrature. The data shown in a and b corresponds to all measurements with total numbers $150<N<200$ and the red lines indicate the inferred standard deviation above and below the mean after subtraction of the readout noise.
		{\bf c, }Classical vs. quantum preparation -- inter-site correlation measurement. Characterization of classical phase noise is crucial to obtain information about the individual quadratures of the signal and idler modes. For initial states prepared classically, that is by linear coupling, strong inter-site correlations between different sites in the optical lattice after mixing with the local oscillator (illustrations) reveal classically washed out coherences (top panel). In contrast, when the ($F=2,m_F=\pm1$) states are populated by spin changing collisions no correlations are observed within the statistical uncertainty showing that $\langle \widetilde{X}_{\uparrow(\downarrow)} \rangle =0$ as a direct result of the spin dynamics. } 
\label{fig3} 
	\end{center}
\end{figure*}

The observable populations do not suffice to characterize the two-mode state; we must access coherences between the modes, the two-mode quadratures. We employ atomic homodyning by analogy to the  experimental technique used in optics, whereby the quadratures are measured relative to a reference field. Yet in the atomic case the population of available local oscillator (LO) reference fields is often limited to rather small atom numbers. This has important consequences for the observable quadratures $\widetilde{X}_{k} = (a_{k}^\dagger a_{\rm LO} + a_{k} a_{\rm LO}^\dagger)/\sqrt{\langle a_{\rm LO}^\dagger a_{\rm LO}\rangle}$, which in general differ from the canonical ones. A measure for this difference is given by the ratio of the quantum mode population to the local oscillator population, by which the Heisenberg uncertainty relation is modified $\Delta \widetilde{X}_{k} \Delta  \widetilde{Y}_{k} \geq 1-{\langle N_{k} \rangle}/{\langle a_{\rm LO}^\dagger a_{\rm LO}\rangle}$~\cite{Ferris:2008}.  \\
For the twin-atom state one expects peculiar coherence properties: The quadratures of the individual modes fluctuate strongly around a vanishing mean, while the two-mode quadratures $X_{\pm}(\varphi)$ show reduced fluctuations for certain $\varphi$ reflecting the quantum correlations. To reveal these characteristics in the coherences of the generated quantum fields we employ a homodyning scheme such that information about the two-mode quadratures $\widetilde{X}_{\pm}(\varphi)$ can be obtained. Measurement of these observables requires comparison of each mode with a local oscillator. We ensure relative local oscillator phase stability by using the pump mode $(2,0)$ as the simultaneous reference, whose phase $\varphi$ can be controlled by microwave dressing. An atomic three-port beam splitter is realized by radio-frequency coupling of signal and idler modes to the pump. Analysis of the fluctuations in atom number sum and difference $N_{(2,-1)} \pm N_{(2,1)}$ between the $(2,-1) $ and $(2,1)$ Zeeman states after the mixing provides, after proper normalization, an upper bound (ub) for the variance of the two-mode quadratures $\Delta^2 \widetilde{X}^{\rm ub}_{\pm}(\varphi)$. The measured fluctuations include further noise contributions governed by population fluctuations in the signal and idler modes, which are small only for the quadrature difference. Further details about the analysis in the three mode framework and the choice of the normalizations  can be found in the supplementary information.
Figures~\ref{fig3}a and b show raw data in the regime of small pump depletion of the measured normalized atom number difference and sum versus local oscillator phase $\varphi$. No phase dependence is visible in the mean, nevertheless, the fluctuations are strongly modulated indicating phase correlations between the two atomic modes.

 We set up a different experiment to obtain the quadrature fluctuations of the signal and idler modes individually. As local oscillators we prepare population in the $(1,\mp1)$ states before initiating the dynamics. Mixing of the local oscillators with the quantum modes is done by a two-photon microwave and radio-frequency $\tfrac{\pi}{2}$ coupling pulse. For both, the signal and idler mode, we find coupling phase independent fluctuations of the atomic quadratures $\widetilde{X}_{\uparrow (\downarrow)} = (N_{(2,\pm 1)}-N_{(1,\mp 1)})/\sqrt{\langle N_{\rm LO}\rangle}$ and a vanishing mean. The mean local oscillator population $\langle N_{\rm LO}\rangle$ of the $(1,\mp1)$ state was measured by omitting the mixing pulse. \\
To assure that our observations are indeed resulting directly from the spin changing collision process the technical phase noise has to be characterized. The 1D optical lattice setup allows for this characterization by a reference measurement in which the population in the $(2, \pm1)$ states is also prepared by electromagnetic coupling and spin changing collisions are tuned off-resonance during the evolution time.  The inter-site correlations observed after the mixing are shown in figure~\ref{fig3}c for this reference experiment (top) and  for the quantum state prepared by spin changing collisions (bottom). In both cases we observe similarly large on-site fluctuations in the atom number difference. However, in the reference experiment we can use the inter-site correlations to reduce these fluctuations by a factor of $20$ such that the remaining fluctuations are close to the expected noise limit for two coherent modes. This shows that inter-site correlations detect finite coherences that have been washed out by classical phase noise in our measurements.
In case of spin changing collisions no correlations are observed within the statistical uncertainty, meaning that the vanishing coherences are indeed caused by the process itself. Given this result, technical phase noise does not influence the sum of the orthogonal quadrature variances $\Delta^2 \widetilde{X}_{\uparrow (\downarrow)} + \Delta^2 \widetilde{Y}_{\uparrow (\downarrow)} = 2 \Delta^2 \widetilde{X}_{\uparrow (\downarrow)}$ (the measured quadrature variance is LO phase independent), such that it is a useful observable to characterize the quadrature fluctuations.
For further details on the inter-site correlation measurement we refer to the supplementary information.

\begin{figure*}[t]
	\begin{center}
		\includegraphics{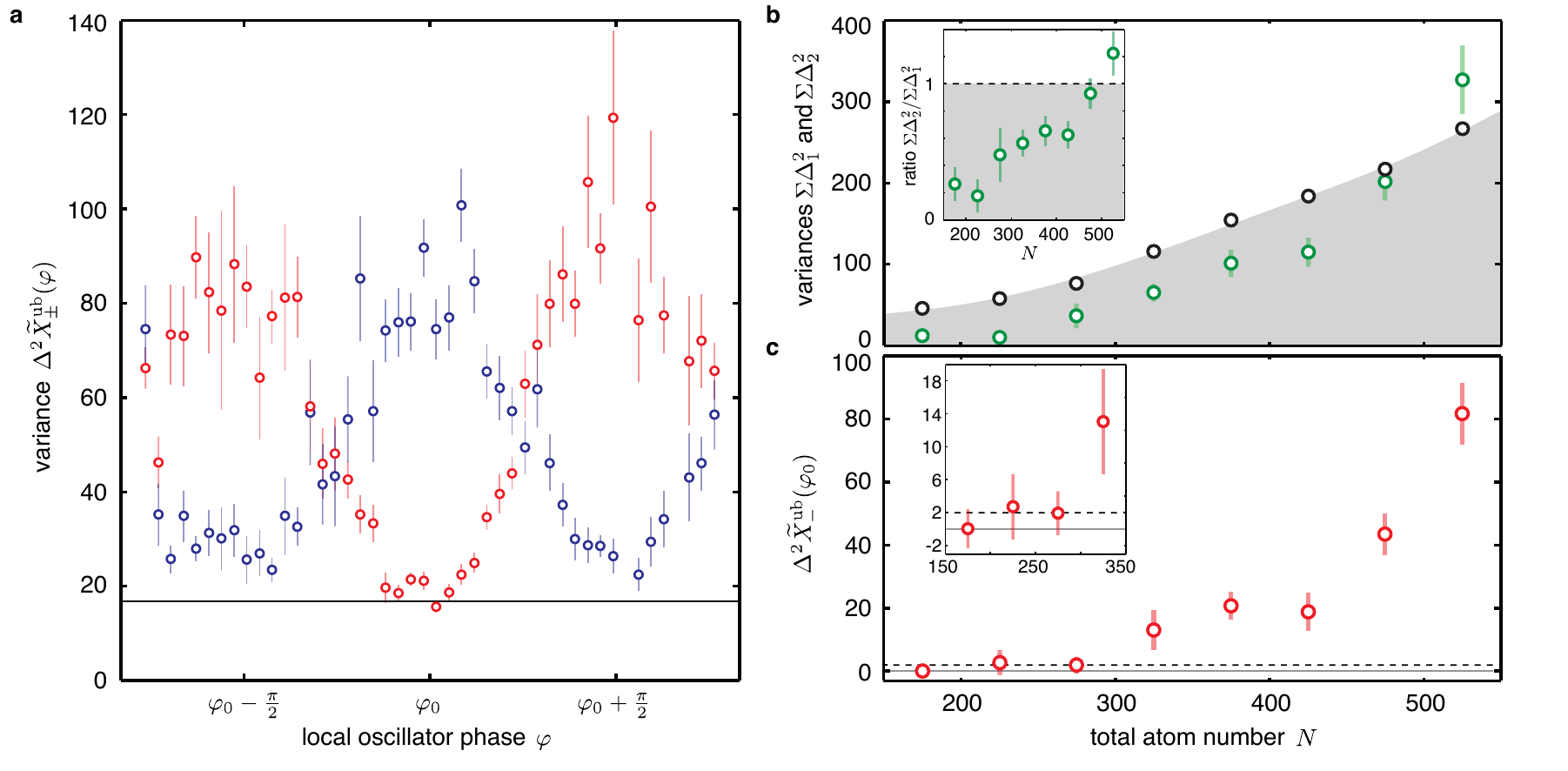} \caption{  {\bf Two-mode quadrature fluctuations and mode inseparability. a, }  Two-mode sum and difference quadratures. The measured variances  $\Delta^2 \widetilde{X}^{\rm ub}_{+}(\varphi)$ (blue) and $\Delta^2 \widetilde{X}^{\rm ub}_{-}(\varphi) $ (red) -- upper bounds for the sum and difference two-mode quadrature variances -- are plotted versus local oscillator phase for total atom numbers $150<N<200$. The shown experimental results are not corrected for readout noise (in contrast to the values stated throughout the text and in b, c) but its magnitude is shown as the black line.
{\bf b, } Continuous variable quadrature entanglement. Inseparability is detected in the gray shaded region where $ \Sigma \Delta_2^2$, the upper bound for the sum of the two-mode quadrature variance minima (green), is smaller than  $\Sigma\Delta_1^2$, the sum of the variance in the individual quadratures (black). The inset shows the ratio $\Sigma \Delta_2^2 / \Sigma \Delta_1^2$.
{\bf c, }  Minimum of the two-mode difference quadrature fluctuations. For small total atom numbers $N$ -- in the non-depleted pump regime and where spin relaxation loss is small -- the minimum two-mode quadrature variance $\Delta^2 \widetilde{X}^{\rm ub}_{-}(\varphi_0)$ inferred from a local quadratic fit is at the noise limit for two coherent fields (dashed black line). The inset shows a zoom into the small $N$ region. All error bars correspond to 1-s.d. uncertainties.  }
\label{fig4} 
	\end{center}
\end{figure*}

Inter-mode entanglement of the twin-atom state is reflected in suppressed noise in the two-mode quadratures compared to the quadrature fluctuations of the individual modes $\Delta^2 \widetilde{X}_- + \Delta^2\widetilde{Y}_+ < \Delta^2{\widetilde{X}}_\uparrow + \Delta^2{\widetilde{X}}_\downarrow + \Delta^2{\widetilde{Y}}_\uparrow + \Delta^2{\widetilde{Y}}_\downarrow$~\cite{Raymer:2003}. When investigating this inequality in our experiment,  care has to be taken to assure comparability of the single- and two-mode measurements~\cite{Ferris:2008}. In both measurements the ratio of the sum of the population in the signal and idler modes to the reference mode population is below $10\%$ for small $N<200$ and grows to approximately $55\%$ for the largest $N$. 
In figure~\ref{fig4}a we plot the measured upper bound for the two-mode quadrature variances versus local oscillator phase $\Delta^2 \widetilde{X}^{\rm ub}_{\pm}(\varphi)$ for an exemplary total atom number. The sum of the minima at $\varphi_0$ and $\varphi_0 - \tfrac{\pi}{2}$ provides an upper bound for the sum of the orthogonal two-mode quadrature variances $\Sigma\Delta_2^2 = \Delta^2 \widetilde{X}^{\rm ub}_{-} + \Delta^2 \widetilde{Y}^{\rm ub}_{+}$, while the measurement of the single-mode quadrature variances $\Sigma\Delta_1^2 = 2 (\Delta^2 \widetilde{X}_{\uparrow} + \Delta^2 \widetilde{X}_{\downarrow})$ has been discussed above. The generated quantum states fulfill the inequality $\Sigma\Delta_2^2 < \Sigma\Delta_1^2$ for a wide range of total atom numbers (Fig.~\ref{fig4}b) showing that the produced twin-atom state is inseparable -- the observed minimal ratio  is $ \Sigma \Delta_2^2 / \Sigma \Delta_1^2 \approx 0.2$.  \\
For the  squeezed vacuum, which is theoretically expected in the non-depleted pump regime, the two-mode quadrature variances should be reduced below the noise level of two coherent modes $\Delta^2 \widetilde{X}^{\rm ub}_{-}(\varphi) = 2$.  For the more precisely detectable variance of the quadrature difference we find the noise minimum for small total $N$ comparable to this level, limited by the experimental signal to noise ratio. In figure~\ref{fig4}c we show the dependence of this noise minimum on the total atom number $N$ -- detection noise subtraction is crucial here and without it we find $\Delta^2 \widetilde{X}^{\rm ub}_{-}(\varphi) \approx 17$ as the minimal value.

In conclusion, we have developed an atomic homodyne detection method, which allows for the measurement of quadrature correlations of twin-atom quantum fields. The continuous variable entangled states have been produced in a deterministic manner utilizing controlled atomic spin interactions -- for small total atom numbers the observed populations distributions are compatible with the atomic two-mode squeezed vacuum. 
Einstein-Podolsky-Rosen (EPR) entanglement  is revealed by $\Delta^2 (X_\uparrow-X_\downarrow) \Delta^2(Y_\uparrow+Y_\downarrow) < 1$ and we find $4\pm17$ for this value, showing that EPR-entanglement for atomic quadratures is within reach. In a future extension of our scheme the local oscillator might be split before the mixing, enabling precise detection also of the two-mode quadrature sum~\cite{Ferris:2008}.   
Spatial separation of the modes can be implemented by employing the Stern-Gerlach technique, being an alternative approach to recently reported experiments~\cite{Jaskula:2010, Bucker:2011}. 
The regime of large average pair population is accessible such that highly non-classical quantum states might be generated by controlled quantum spin evolution~\cite{Diener:2006}.

We note that in parallel to this work, sub-poissonian number fluctuations after spin changing collisions have been observed by two other groups~\cite{Bookjans:2011, Lucke:2011} and it has been shown that the generated quantum correlated states are useful for noise interferometry beyond the standard quantum limit~\cite{Lucke:2011}. 

\newpage

\appendix
\section{Supplementary Information}
\subsection{Experimental sequence}
We routinely produce Bose-Einstein condensates (BEC) of Rubidium 87 in the low field seeking $(F, m_{F})=(1,-1)$ hyperfine state with an experimental cycle time of approximately $40\,$s. Before the final evaporation ramp in an optical trap we turn up a 1-dimensional optical lattice slicing the atomic cloud into eight pieces. Since the lattice potential does not allow for tunneling between different sites on the experimental timescale, we start with eight independent samples with different mean total atom number $N$ per site on which the spin dynamics experiment is done in parallel. \\
First we transfer the atoms to the $(1,0)$ state by a  radio frequency (rf)  $\pi$ pulse. A homogeneous magnetic offset field of approximately $9\,$G is applied such that the second-order Zeeman shift is sufficient to resolve the different rf transitions in the $F=1$ hyperfine multiplet.  Afterwards we remove spurious population in the $(1,\pm1)$ states by a Stern-Gerlach technique. Care is taken to adiabatically ramp down the offset field to its final value of $B_{0}\approx 1.5\,$G where we actively stabilize it by a fluxgate sensor based feedback loop~\cite{Gross:2010}. The offset field is chosen such that both, state selective two-photon pulses between the $(1,-1)$ and the $(2,1)$ states, as well as power broadened symmetric rf coupling between the $(2,0)$ and the $(2,\pm1)$ states are possible. For the homodyning of single-mode quadratures local oscillators in $(1,\pm1)$ are necessary, which we prepare by transferring a part of the population via symmetric rf coupling. The next step is a microwave $\pi$ pulse from $(1,0)$ to $(2,0)$. \\
Spin dynamics is initiated by compensating the second-order Zeeman and mean field shift (in total $\delta = 162\,$Hz) with microwave dressing of the $(2,0)$ level~\cite{Kronjaeger:2006,Gerbier:2006}.
The dressing field is $98\,$kHz blue detuned to the $(1,0) \leftrightarrow (2,0)$ transition with a resonant microwave Rabi frequency of $8\,$kHz. This selective technique has the advantage that only spin dynamics between the $m_{F}=0,\pm1$ states is resonant. In all experiments we allow for $22\,$ms of spin evolution in which the $m_{F}=\pm1$ states become populated via spin changing collisions. \\
For the measurement of the single-mode quadratures we now apply a two-photon $\pi/2$ pulse resonant with the $(1,\mp1)  \leftrightarrow (2,\pm1)$
transitions, where changing the phase of these pulses is equivalent to a change of the local oscillator phase.\\
In order to realize the three-port beamsplitter scheme, i.e. to measure the two-mode quadratures after the spin evolution, we apply a single rf field for $60\,\mu$s, which couples the $(2,0)$ and $(2,\pm1)$ states. This corresponds to a $\pi/5$ coupling pulse, where the pulse angle is restricted by experimental considerations described below. For control of the local oscillator phase  $\varphi$ (of the $m_{F}=0$ state) we switch off the dressing field for a variable time $t$ (between $0\,$ms and $4.5\,$ms) before the coupling pulse, such that $\varphi = 2\pi \delta t$.
In the case of the measurement of population difference $N_{-}$ no coupling is applied after the spin changing collisions. \\
After the experimental sequences described above the population in the different Zeeman states is measured.
State and site selective imaging is achieved by a combination of the Stern-Gerlach technique and a high spatial resolution imaging system~\cite{Gross:2010,Esteve:2008}. The Stern-Gerlach pulse, which separates the $m_{F}$ states, is aligned carefully with the magnetic offset field such that the following absorption imaging is a projective measurement of the population in the Zeeman states $m_F$ defined with the quantization axis along the direction of the offset field.

\begin{figure}[t]
\begin{center}
\includegraphics{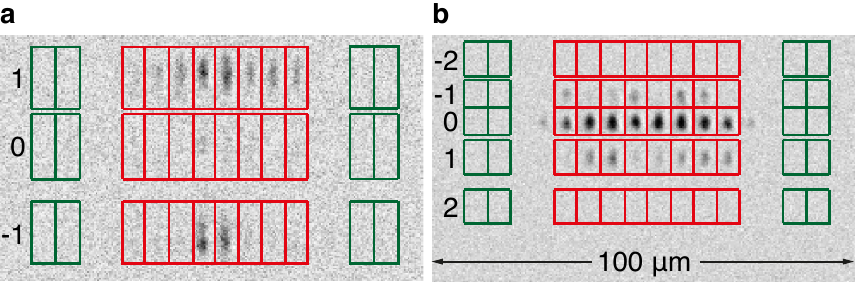}
\caption{
{\bf Experimental raw data. } Absorption picture of atoms in the $F=1$ ({\bf a}) and $F=2$ ({\bf b})  multiplet. The counting boxes for regions with atomic signal are shown in red, the boxes for technical noise measurement in green. The numbers indicate the Zeeman sub state, which is the same in each row. Due to the opposite sign of the magnetic moments these numbers are inverted between $F=1$ and $F=2$. The size of the image in real space is given. Note that the clouds for $F=1$ atoms are more extended in vertical direction due to the imaging sequence~\cite{Gross:2010}.}
\label{figRawPic}
\end{center}
\end{figure}

\begin{center}
\begin{figure*}[ht]
\includegraphics{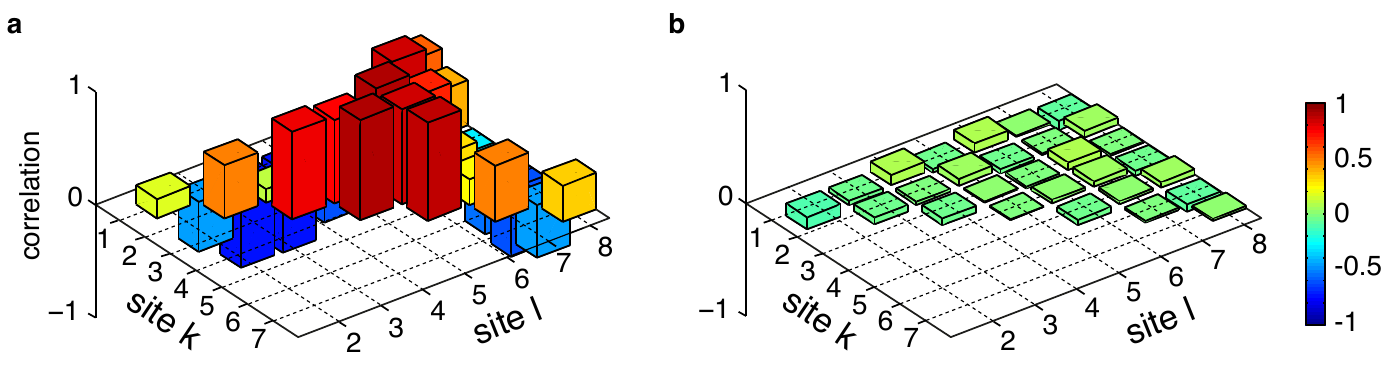}
\caption{
{\bf Experimentally observed inter-site correlations. a,} Initially coherent state. Strong inter-site correlations $C_{kl}$ are detected for an initially coherent state, which are caused by spatially large scale technical fluctuations resulting in site correlated phase-noise. {\bf b,} Spin-changing collisions experiment. After spin dynamics a random phase is observed and the inter-site correlations vanish within the statistical uncertainty $C_{kl}\approx 0$. The striking difference in the correlation signal between the two measurements shows that environmental noise can neither explain the random phase of the individual modes nor the correlations between signal and idler modes observed after spin-changing collisions.}
\label{figCorr}
\end{figure*}
\end{center}

\subsection{Data analysis}
Imaging is done in the high intensity regime and the column density (atom number per pixel) is calculated from the raw absorption data following the recipe in~\cite{Reinaudi:2007}. Care is taken to assure a linear and well calibrated imaging system~\cite{Gross:2010,Esteve:2008}. To obtain the atom number per state and lattice site we define counting boxes in which the column density is integrated. Figure~\ref{figRawPic} shows a typical single run picture with counting boxes indicated. Two types of technical noise add to the atomic signal in the imaging process. The first is unavoidable photon shot noise ($\sqrt{\Delta^2 N^{\rm psn}} \approx 7.0$ atoms per box) which we deduce from the data in each experimental run~\cite{Gross:2010,Esteve:2008}. However, interference fringes due to the coherence of the imaging light make absorption imaging very sensitive to mechanical vibrations and cause excess noise. Even though we discard images which show obviously strong interference fringes in the regions with no atomic signal (one picture out of $50$), the remaining ones still have a finite noise contribution due to large scale interference structures. These are very hard to avoid, but on an ensemble of pictures their contribution to the detected variance can be well measured. Hence we define counting boxes in regions without atoms of the same size as those used to extract the atom numbers (see Fig.~\ref{figRawPic}) and measure the excess variance. It corresponds typically to a noise level of $\sqrt{\Delta^2 N^{\rm fr}} \approx 3.9$ atoms per box where we checked for the spatial homogeneity of the fringe noise over the relevant regions of the picture. 
The technical noise is gaussian and not correlated with the atomic noise to be measured. Therefore we can subtract the independently measured noise contributions from the measured atom number variances. Note that this subtraction is crucial for the reported results since the total background noise level corresponds to a variance of $\Delta^2N_-^{\rm tech} \approx 125$. Without the noise correction we would measure sub-shot-noise fluctuations only for $N \gtrsim 450$ where $\langle {N}_+ \rangle > \Delta^2N_-^{\rm tech} $ (see figure 2 of the main text). Here it is important to note that the
uncertainties on the experimental data given in this manuscript include the statistical uncertainty due to the noise subtraction. As a test for our noise calibration we analyzed pictures without atoms in the counting boxes and found possible systematic errors smaller than the statistical uncertainties.  \\
The dependence of all presented observables on the total atom number $N$ is obtained by binning the data in intervals of size $\delta N = 50$.

\subsection{Inter-site correlation analysis}
The use of a single site resolved optical lattice in our experiments has various advantages. Two of them are rather obvious, the boost in statistics due to the parallel realization of eight experiments and the increased local confinement which is important for the validity of the single mode approximation. Here we point out that inter-site correlations between the observables can be used to extract phase noise contributions stemming from environmental fluctuations. \\
At a finite magnetic field $B$ the energy of the different Zeeman states shifts differentially with $B$ which leads to phase noise in Ramsey type experiments~\cite{Gross:2010}, in which the integrated phase difference is mapped onto an observable population difference. The homodyning experiments reported in this manuscript are of similar type and the observed atom number differences are in principle sensitive to environmentally induced phase noise.  Magnetic field or microwave phase fluctuations in our experiment are mainly low frequency and homogeneous over a spatial region much bigger than the size of the entire lattice system. Therefore they result in shot-to-shot fluctuations which are correlated between different lattice sites $k$. \\
Figure~\ref{figCorr}a shows the measured inter-site correlations for a reference homodyning experiment with the local oscillator in the $(1,-1)$ Zeeman state and a coherently prepared state in the $(2,1)$ Zeeman state.
The inter-site correlations $C_{kl}$ are calculated for the observed atom number differences $n_k=N_{1,k}-N_{2,k}$ ($1$ and $2$ label the two Zeeman states involved)
\begin{equation}
	C_{kl} = \frac{\langle \widetilde{n}_k \widetilde{n}_l \rangle}{\Delta n_k \Delta n_l}
\end{equation}
where $\widetilde{n}_k = n_k - \langle n_k \rangle$ is the atom number difference in site $k$ corrected for its mean and $\Delta n_k$ the standard deviation of $n_k$. The experimental sequence is similar to the spin-changing collision evolution described above but spin dynamics was suppressed by tuning the process off resonance. Since the relative phases are in this case initially defined by the coherent preparation (corresponding to a non-vanishing mean of the single-mode quadrature), finite inter-site correlations $|C_{kl}|>0$ are observed. Their strength and sign depend on the difference in the total population per site $N_k - N_l$ which is due to the $N_k$ dependent mean field shift. In our experiment $N_k$ fluctuates little from shot to shot. Thus, this population dependent energy offset results in a fixed difference in the integrated relative phases in different sites which causes correlations or anti-correlations depending on its magnitude. Using the inter-site correlations to remove these technically induced phase fluctuations we reduce the observed fluctuations $\tfrac{\Delta n_k^2}{\langle N_{1,k}+N_{2,k} \rangle}$ from approximately $50$ to $2.5$ times the noise limit of two coherent states. This shows that almost all environmental phase noise in our experiment is of common mode type and can be detected by the inter-site correlations.\\
In contrast, for a quantum state with initially vanishing coherences such inter-site correlations cannot build up such that one expects $C_{kl} \approx 0$. 
In figure~\ref{figCorr}b we plot the inter-site correlations for the signal mode homodyning experiment. Here the $(2, 1)$ Zeeman state was populated by spin evolution.  
No statistically significant correlations are  observed, excluding that the phase is randomized by environmental noise. We also tested for inter-site correlations in the signal-idler mode entanglement measurements (Fig.~3a, b and 4 of the main text) and found a vanishing signal $C_{kl} \approx 0$ excluding a significant influence of environmental noise. \\
We emphasize that this characterization of classical de-phasing is crucial for the results presented in figure~4b of the main text, since  we use different local oscillators to deduce the single- and two-mode quadratures. Stronger classical de-phasing in the single-mode quadrature measurement compared to the two-mode measurement would result in a false bound for two-mode inseparability.

\subsection{Two-mode quadratures}
The two-mode quadratures of signal (${a}_{\uparrow}$) and idler (${a}_{\downarrow}$) modes are measured by  simultaneous symmetric radio-frequency coupling to the pump mode ${a}_{0}$ which serves as the local oscillator. This process is described by a unitary rotation ${U}_{\rm cpl}={\rm e}^{-i {H}_{\rm cpl} t/\hbar}$ in the three mode system generated by the Hamiltonian
\begin{equation}
{H}_{\rm cpl} = \frac{\hbar \Omega}{2 \sqrt{2}} ( {a}_{0} {a}_{\downarrow}^{\dagger} +  {a}_{0} ^{\dagger}{a}_{\uparrow}   + {a}_{0} ^{\dagger}{a}_{\downarrow} +  {a}_{0}{a}_{\uparrow}^{\dagger}  ) \label{eq.Hcpl}
\end{equation} 
where $2\pi\hbar$ is Planck's constant.

Experimentally the coupling is switched on and off non-adiabatically with a total pulse duration $\tau_{\rm cpl}$ . The rotation (mixing) angle of the pulse is given by $\gamma=\Omega \, \tau_{\rm cpl}$, which is defined such that a $\gamma = \pi$ pulse transfers all population from the $m_{F}=0$ state to the $m_{F}=\pm1$ states (assuming they are initially empty).   \\
We choose a mixing angle of $\gamma \approx \pi/5 $, 
since we have to make a tradeoff between signal-to-noise ratio (optimal for $\gamma=\pi$) and an unwanted spurious coupling between the $m_{F}=\pm 1$ and $m_{F} = \pm 2$ states, favoring a smaller mixing angle $\gamma$.

The unitary transformation generated by the coupling Hamiltonian (\ref{eq.Hcpl}) is described by the matrix 
\begin{equation}
U = 
\begin{pmatrix}
\frac{\Co+1}{2} & -\frac{i \Si}{\sqrt{2}} & \frac{\Co-1}{2}\\
-\frac{i \Si}{\sqrt{2}} & \Co & -\frac{i \Si}{\sqrt{2}}\\
\frac{\Co-1}{2} & -\frac{i\Si}{\sqrt{2}} & \frac{\Co+1}{2}
\end{pmatrix}
\end{equation}
with $\Si=\sin(\tfrac{\gamma}{2})$, $\Co=\cos(\tfrac{\gamma}{2})$. The mode operators  $({a}_\downarrow', {a}_0', {a}_\uparrow')^T$ in the Heisenberg-picture after the coupling pulse are obtained by applying the matrix $U$ to $({a}_\downarrow, {a}_0, {a}_\uparrow)^T$. The phase shift $\varphi$ of the local oscillator before the pulse is described by the substitution $a_0\rightarrow \tilde{a}_0(\varphi) = a_0{\rm e}^{i\varphi}$. For the sake of clarity this abbreviation is used in the following without the explicit phase dependence.

\subsubsection{Two-mode quadrature difference}
For the experimentally measured population difference $N_-' = {a}_\downarrow'^\dagger {a}_\downarrow'- {a}_\uparrow'^\dagger {a}_\uparrow'$ in the $m_{F}=\pm1$ states we find
\begin{eqnarray}
    N_-'(\varphi) &=& \Co N_-     \nonumber  \\   
            &+& \frac{i \Si}{\sqrt{2}} ( {a}_\downarrow \tilde{a}_{0}^\dagger  +  {a}_\uparrow^\dagger \tilde{a}_{0}  -{a}_\downarrow^\dagger \tilde{a}_{0}   -   {a}_\uparrow \tilde{a}_{0}^\dagger)          
\end{eqnarray}
The second term is, up to a $\tfrac{\pi}{2}$ phase shift of the ${a}_0$ mode, proportional to the generalized two-mode quadratures
\begin{eqnarray}
\widetilde{X}_{-}(\varphi) &=& ( {a}_\downarrow \tilde{a}_{0}^\dagger  -  {a}_\uparrow^\dagger \tilde{a}_{0}  + {\rm h.c.})/\sqrt{\langle \tilde{a}_{0}^\dagger \tilde{a}_{0} \rangle} \nonumber \\
&=& \widetilde{X}_{\downarrow}(\varphi) - \widetilde{X}_{\uparrow}(\varphi)  
\label{eq.gen_quad_diff}
\end{eqnarray}
In the limit of a large local oscillator ($\tilde{a}_0 \approx \sqrt{\langle N_0\rangle}{\rm e}^{i\varphi}$) the generalized two-mode quadratures correspond to the canonical ones~\cite{Ferris:2008}. 
With the normalization factor $\langle N_0 \rangle \Si^2  / 2$ and by using $\langle N_-\rangle=0 $ and $\Ta = \tan(\tfrac{\gamma}{2})$ we obtain the measured upper bound
\begin{eqnarray}
    \Delta^2 \widetilde{X}^{\rm ub}_-(\varphi) 
    &=& \Delta^2 {N}_-'(\varphi)/(\langle N_0 \rangle \Si^2  / 2)\\ 
    &=& \Delta^2 \widetilde{X}_-(\varphi)  \nonumber  \\ 
    &+& \frac{2}{\Ta^2 \langle N_0 \rangle} \Delta^2 {N}_-     \nonumber  \\   
    &+& \frac{\sqrt{2}}{\Ta \sqrt{\langle N_0 \rangle}}  
    \langle N_-  \widetilde{X}_-(\varphi) +  \widetilde{X}_-(\varphi) N_- \rangle   \nonumber                            
\end{eqnarray}
The symbol $\Delta^2$ indicates the variance of the subsequent variable.
The local oscillator phase independent variance offset (second term) which is proportional to the variance of the population difference $\Delta^2 {N}_-$ is small (approximately $0.1$) for small total atom numbers ($150<N<200$) and increases to approximately $3.2$ for $500<N<550$. 
The terms in the last line vanish due to their strong sensitivity to magnetic field fluctuations ($700\,$Hz/mG). At this sensitivity the coherence time in our experiment is approximately $1.5\,$ms which results in $\langle N_-  \widetilde{X}_- \rangle \approx 0$ after $22\,$ms of spin evolution.\\

\subsubsection{Two-mode quadrature sum}
The second important two-mode quadrature
\begin{eqnarray}
\widetilde{X}_{+}(\varphi)&=& ( {a}_\downarrow \tilde{a}_{0}^\dagger  +  {a}_\uparrow^\dagger \tilde{a}_{0}  + {\rm h.c.})/\sqrt{\langle \tilde{a}_{0}^\dagger \tilde{a}_{0} \rangle} \nonumber \\
&=& \widetilde{X}_{\downarrow}(\varphi) + \widetilde{X}_{\uparrow}(\varphi),
\end{eqnarray}
i.e., the sum of the two single mode quadratures, can be measured less precise in the current scheme, however, an upper bound for its variance can be obtained. For the sum of the population in the $m_{F}=\pm1$ states $N_+' = {a}_\downarrow'^\dagger {a}_\downarrow'+ {a}_\uparrow'^\dagger {a}_\uparrow'$ we obtain
\begin{eqnarray}
    N_+' &=& \tfrac{\Co^2+1}{2} N_+ + \Si^2 {N}_{0} \nonumber \\
	     &+& \tfrac{i\Co\Si}{\sqrt{2}} ({a}_\downarrow \tilde{a}_{0}^\dagger - {a}_\uparrow^\dagger  \tilde{a}_{0} -  {a}_\downarrow^\dagger \tilde{a}_{0}   + {a}_\uparrow\tilde{a}_{0}^\dagger  )  \nonumber \\
		&+& \tfrac{\Co^2-1}{2} ({a}_\downarrow^\dagger {a}_\uparrow + {a}_\uparrow^\dagger {a}_\downarrow)
		\label{eq.sumquad}
\end{eqnarray}
The third term $\tfrac{i \Co\Si}{\sqrt{2}} ({a}_\downarrow\tilde{a}_{0}^\dagger - {a}_\uparrow^\dagger \tilde{a}_{0} -  {a}_\downarrow^\dagger \tilde{a}_{0}   + {a}_\uparrow\tilde{a}_{0}^\dagger  )$ is proportional to the two-mode quadrature $\widetilde{X}_{+}(\varphi+\tfrac{\pi}{2})$, which is the only local oscillator phase dependent term. Covariance terms between this one and the three other terms in equation (\ref{eq.sumquad}) vanish due to their strong sensitivity to magnetic field fluctuations (see above). Hence, one obtains the upper bound for the two-mode quadrature sum
\begin{eqnarray}
	\Delta^2 \widetilde{X}_+^{\rm ub}(\varphi) 
	&=& \Delta^2 {N}'_+(\varphi)/(\langle N_0 \rangle \Co^2\Si^2  / 2)\nonumber\\
	&=& \Delta^2 \widetilde{X}_+(\varphi) + {\rm const.}
\end{eqnarray}
where the constant offset due to the variance of the phase independent terms in equation (\ref{eq.sumquad}) limits the obtainable precision when deducing $\Delta^2\widetilde{X}_+(\varphi) \leq \Delta^2{N}'_+(\varphi)$  from our measurements. Note that the normalisations for $\widetilde{X}_+$ and $\widetilde{X}_-$ are different.
A future extension of this scheme to measure also the sum of the single mode quadratures with high precision and therefore both quadrature Einstein-Podolsky-Rosen variables~\cite{Reid:2009} would require the splitting of the local oscillator prior to the coupling, such that the phase of the two coupling fields can be tuned independently~\cite{Ferris:2009}.

\begin{acknowledgments}
 We acknowledge enlightening and clarifying discussions with P. Grangier, A. Aspect, A.J. Ferris, M.J. Davis and B.C. Sanders. This work was supported by the Forschergruppe FOR760, Deutsche Forschungsgemeinschaft, the German-Israeli Foundation, the Heidelberg Center for Quantum Dynamics, Landesstiftung Baden-W\"{u}rttem\-berg, the ExtreMe Matter Institute and the European Commission Future and Emerging Technologies Open Scheme project MIDAS (Macroscopic Interference Devices for Atomic and Solid-State Systems). G.K. acknowledges support from the Humboldt-Meitner Award and the Deutsche-Israelische Projektgruppe (DIP). \\
 Correspondence should be addressed to M.K.O. (homodyning@matterwave.de).
\end{acknowledgments}

\end{document}